# Fast coordinate cross-match tool for large astronomical catalogue

Vladimir Akhmetov
*Laboratory of Astrometry*
*Institute of Astronomy,*
*V. N. Karazin Kharkiv National University*
Kharkiv, Ukraine
akhmetovvs@gmail.com

Artem Dmytrenko
*Department of Astronomy and Space Informatics,*
*V. N. Karazin Kharkiv National University*
Kharkiv, Ukraine
astronom.karazin007@gmail.com

Sergii Khlamov
*Laboratory of Astrometry*
*Institute of Astronomy,*
*V. N. Karazin Kharkiv National University*
Kharkiv, Ukraine
sergii.khlamov@gmail.com

*Abstract*—In this paper we presented the algorithm designed to efficient coordinate cross-match of objects in the modern massive astronomical catalogues. Preliminary data sort in the existed catalogues provides the opportunity for coordinate identification of the objects without any constraints with the storage and technical environment (PC). Using the multi-threading of the modern computing processors allows speeding up the program up to read-write data to the storage. Also the paper contains the main difficulties of implementing of the algorithm, as well as their possible solutions.

## I. INTRODUCTION

In recent years in astronomy the development of telescope- and instrument-making has led to an exponential growth of the observational data. The modern astronomical catalogues are the 2D-spreadsheets that contain the various information about the celestial bodies. An each row of this table corresponds to the data of one object.

There are lot of information about the object in this row, such as:

- position in spherical coordinate system;
- errors in determining of the coordinates;
- stellar magnitude in the different photometric bands (brightness of the object);
- standard errors of stellar magnitude;
- proper motions and other useful information.

The number of objects in the modern astronomical catalogues reaches to the several billion objects, and the size of tables that contain information about these objects varies from hundreds of GB to the several TB.

So, the knowledge extraction from such data will be the most complicated challenge for researchers and scientists.

In the near future the following telescopes will be launched: Large Synoptic Survey Telescope (LSST) [1] (Fig. 1) and Thirty Meter Telescope (TMT) [2] (Fig. 2). Both telescopes will give about 30 TB of data for one observational night.

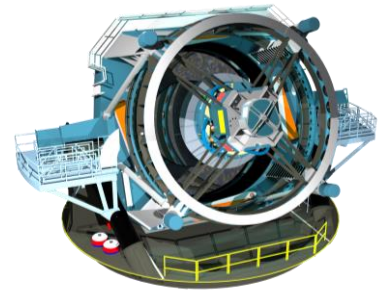

Fig. 1. Large Synoptic Survey Telescope (LSST).

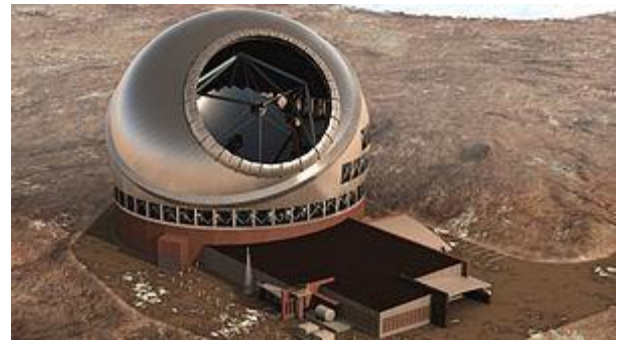

Fig. 2. Thirty Meter Telescope (TMT).

Even today, the scientists of the world are faced with the problem of the large data during the following missions:

- ESA GAIA space mission [3, 4]: 3D-map of Milky Way with collecting of about 1 PB of data in 5 years for 1.3 billion objects;
- Pan-STARRS [5]: collecting of more than 100 TB of data for more than 2 billion objects;
- ESA Euclid space mission: collecting of more than 200 TB of data (less than 800 GB/day over at least 6 years).

The data mining techniques and intelligent management technologies of data analysis are rapidly evolving, but cross-matching is still one of the main step of any standard modern pipeline for data analysis or reduction.

For example, a pipeline from the following software includes estimation of the objects position (data analysis), astrometry and photometry reduction: CoLiTec (Collection Light Technology) software (http://neoastrosoft.com) [6, 7], Astrometrica [8] and others.

One of the main step of comparing and analyzing the data is the coordinate identification of common objects in the modern massive astronomical catalogues, Big Data or any large data sets or streams that contain useful information about celestial objects. For this purpose the different databases are used, but all of them are based on MSSQL for Windows and PostgreSQL for Unix systems.

This approach is very convenient for the storing and obtaining the quick access to data from various tables (catalogues). The such approach allows developing the software for analysis of the data from the different tables (catalogues) of database. Also, there is an opportunity in database for coordinate identification of the objects in the different catalogs.

The example of such database of astronomical catalogues is VizieR (http://vizier.u-strasbg.fr). It is a joint effort of CDS (Centre de Données astronomiques de Strasbourg) and ESA-ESRIN (Information Systems Division). VizieR has been available since 1996, and was described in a paper published in 2000 [9]. VizieR includes 18 282 catalogues that are available from CDS. 17 629 catalogues from all of them are available online as full ASCII or FITS files. 17 342 catalogues are also available through the VizieR browser.

Using the online access to the different astronomical catalogues provided by VizieR [9], the various software, such as CoLiTec [6, 10] and Astrometrica [8], can perform data analysis using different data mining techniques and intelligent management technologies.

In addition to the obvious advantages of using such databases, there are a number of disadvantages that need to be corrected. In this paper we presented the one of available algorithms for the quick coordinate identification of common objects (intersection) in the modern massive astronomical catalogues without using of the implemented to database algorithms.

II. CROSS-IDENTIFICATION

The modern catalogues include values of stellar magnitude in different photometric bands that are also can be obtained at the various epoch. In this case, we could not use photometry for cross-identification of these catalogues, therefore we had to perform cross-identification using only coordinates of objects. Such cross-identification is not necessarily an exact identification.

In this paper we described some important steps for developing of the cross-identification method. It was used for creating catalogue with 421 million positions and absolute proper motions of stars (PMA) [11] and for cross-identification of PMA, PPMXL, UCAC4, UCAC5, HSOY, TYCHO2, and TGAS catalogues [12],[13].

In [11] the windows with different sizes ranging from 0.1 to 15 arsec with a step 0.1 arcsec are used because of a very large difference of stellar density at the different galactic latitudes. We counted the increment of a number of stars dN (Fig. 3, blue points), which fell into the circular zones with radii R and R + dR. This increment is a function of the ring radius and can be represented by a sum of two independent functions (Fig. 3, green points).

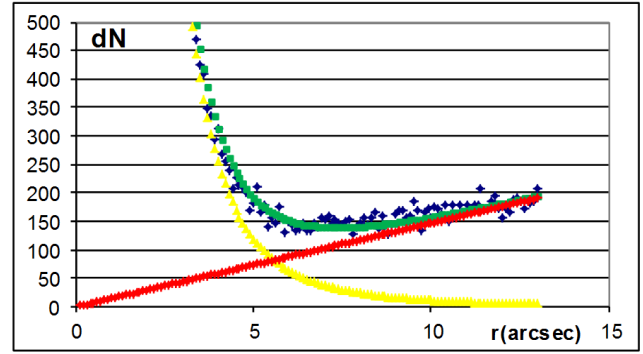

Fig. 3. The increment of a number of stars as a function of the ring radius.

The first function describe uniform density distribution (Fig. 3, red points) of stars over the sky pixel and is directly proportional to the radius of window. The second one is the density distribution function of angular distances for the nearest neighbors. The distribution function can be calculated for the random (Poisson) distribution (Fig. 3, yellow points) of star positions.

The intersection point of these two functions allowed us to establish the optimal window size for cross-identification of catalogues. This point corresponds to such radius where the probability of misidentification reaches the probability of omitting a star with a considerable proper motion.

The described algorithm does not guarantee a correct identification for all objects from the catalogues, but according to our research and analysis the almost all objects have been identified correctly [11].

Let's represent the data of various astronomical catalogues in the form of sets A and B. So, the result of cross-identification of these astronomical catalogues will be represented as one of the combinations of join type in the figure 4.

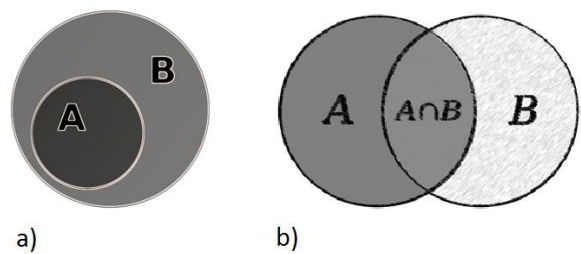

Fig. 4. Combinations of join type.

In the first example (Fig. 4a) A set completely belongs to B set. In this case the result of cross-identification of sets A and B

will be all objects from A set and the number of objects can be predicted.

Usually, the situation with the coordinate identification (intersection) of objects in large astronomical catalogues looks much more complicated. In general, this complication is due to the large random and systematic errors in determining of the objects coordinates. Also the complication can be caused by the significant difference between the epochs of observations and the stellar proper motions.

Because of the described reasons the coordinates of object in various catalogues can be also different. So, the result of intersection of two catalogues will be only objects whose coordinates do not exceed the optimal search radius. This intersection result can be also represented as subset of common objects and as combinations of sets A and B (Fig. 4b).

### III. ASTRONOMICAL DATABASE

During research the special database has been developed. It provides a quick and simple access to the modern astronomical catalogues that contain data of millions or even billions celestial objects including stars, galaxies, quasars and others data.

All of these modern astrometric catalogues were collected in the database using MSSQL (Windows) server. This database contains about 50 catalogues with data more than 2 TB.

The list of some modern astronomical catalogues that were obtained for the last several years is provided in the table 1. The table 1 also contains appropriate size of catalogues and amount of objects that are included in the catalogue.

TABLE I. GENERAL DATA ABOUT CATALOGUES FROM DATABASE

| Catalogue Name | Size, MB | Amount of objects |
|---|---|---|
| GAIA DR1 | 73 173.688 | 1 142 679 769 |
| GPS1 | 38 110.445 | 341 469 435 |
| GSC23 | 42 717.320 | 940 464 379 |
| HSOY | 117 749.281 | 583 001 652 |
| PMA | 33 944.477 | 421 454 398 |
| PPMXL | 184 754.203 | 910 468 710 |
| PSC | 41 814.016 | 470 992 970 |
| TGAS | 131.742 | 2 057 050 |
| UCAC5 | 10 142.953 | 107 758 513 |
| USNOB1 | 98 337.547 | 1 044 738 050 |
| WISE | 100 128.141 | 563 921 584 |
| XPM | 29 167.617 | 313 610 083 |
| XPM2 | 117 749.281 | 1 077 651 504 |

Also the special web-interface has been created to facilitate access to the all available astronomical data of these catalogues. The web-interface was written using PHP programming language and available by the following address: http://astrodata.univer.kharkov.ua/astrometry/db.

The database allows the user to carry out data selection that containing in a small region of the celestial sphere from the large astronomical catalogues by using: a local network, an internet browser, special scripts and programs.

### IV. CROSS-MATCH ASTRONOMICAL CATALOGUE BY MEAN OF DATABASE

The CDS cross-match service is a new tool, which allows astronomers to efficiently cross-identify sources between very large catalogues (up to 1 billion rows) or between a user-uploaded list of positions and a large catalogue (http://cdsxmatch.u-strasbg.fr/xmatch).

In this service you can perform a cross-match based only on the positions lying at an angular distance less than radius $R$ (1), or a cross-match based on positions taking into account error uncertainties.

$$R = 3 * sqrt(eRA_1^2 + eDEC_1^2 + eRA_2^2 + eDEC_2^2), \quad (1)$$

where $eRA_1$, $eDEC_1$ – the uncertainties in right ascension and declination of objects in the first catalogue; $eRA_2$, $eDEC_2$ – the uncertainties in right ascension and declination in the second catalogue.

Cross-match can be performed on all sources of both tables, or can be restricted to a cone around a given position or object name, or to a given HEALPix cell [14].

It turns out that when using CDS cross-match service, in addition to advantages, there are several serious disadvantages that can greatly affect to the results of the intersection. The main disadvantage of this service is the incorrect identification of objects in the areas with high density, where more than one object falls into the search window. The processing result of this service is the selecting of all possible pairs of objects, but the total number of crossed objects may exceed 100% of objects in set A (Fig. 4a).

For example, Tycho2 catalogue contains 2 539 913 objects in total. Cross-match between itself (Tycho2 x Tycho2) should give the result equals to 2 539 913 – the same number of objects in total of Tycho2 catalogue. But when we performed cross-match between itself only for the small search window with size at 1.5 arcsec, the CDS cross-match service has returned the number of common objects equals to 2 543 009.

So, for the catalogues that contain more than billion objects the number of false-identified objects is from 5 to 10 percents depending on the objects density in the area of the sky (the number of objects per one square degree).

### V. FAST CROSS-MATCH OF CATALOGUES WITHOUT USING DATABASE

The cross-match between catalogues that include hundreds of millions of data units is a big technical problem. Mainly because these data cannot be kept in RAM. Of course, the disk storage also should be very large for this purpose.

Usually the input data of the astronomical catalogues are organized in format of tables using the following databases: MSSQL, MySQL or PostgreSQL. But the storage engine is more suitable to the fast reading after carry out of indexing.

The output data of the catalogues cross-match are written to the new table. But as shown the results of testing, actually the data writing to the database rather than the data reading was a bottleneck. In the current research we did not use any database for the cross-match of astronomic catalogues.

All data are stored in a text files that are sorted by declination. The source code of the program was written using C++ language with supporting of the CPU multithreading. The optimization was a compromise between CPU usage and RAM limitations. In our case source code was optimized for the performance on the server with 32 GB RAM, Intel(R) Core(TM) i7 CPU with 4 cores at 3.2 GHz with hyper threading for a total of 8 CPUs.

In the developed method the calculations are performed in RAM and the input data for both the first and the second catalogues are read from previously sorted files by small areas of declination (several millions objects). In addition, the input data of catalogue are divided into several (depend from amount of hyper threading) declination strips and the calculations of objects pairs for the different strips are run in parallel.

The developed tool for cross-match uses C data structures as some "string data" with RA and DEC double type and boolean flag (true or false). This "flag" equals to "true" if the object does not have any pair from the second catalogue. The "flag" equals to "false" when the object is have a pair and cannot used for other pairs calculation. In case, when several objects are in the optimal radius of search, the nearest object will be selected.

The developed method for cross-match of catalogues includes only one object from the area with high density. After cross-match of objects for the current area the result will be presented as two "string data" from the first and second catalogues. This data will be written to the text file as one object and then the tool moves on to the next area and read it and so on.

It should be noted that the time for reading and writing the data result of cross-match more than 70% from total time in case of using 8 CPUs (threading).

## VI. Conclusions

In this paper we developed fast cross-match method for the large astronomical catalogues without using a database. Using C++ programming language we developed the tool for cross-match based on the appropriate method, which was successfully used during our research. Also this tool will be used for the further analysis of the following large astronomical catalogues: GAIA DR1, PMA, WISE, UCAC5, HSOY, GPS1, PPMXL, GSC2.3 and XPM.

With help of the preliminary sorted by declination data and using hyper threading the developed tool can carry out cross-match data from the large astronomical catalogues on the time approximate to the time of read and write the data and processing results from/to HDD.

The developed method excludes an opportunity of multiple cross-match of objects in areas with high density. In this case only one (nearest) object can be presented as a result of cross-match. This approach is very important particularly for the modern massive astronomical catalogues that contain data with more than several billion objects.

ACKNOWLEDGMENT

The authors thank CDS (Strasbourg, France) who provided online access to the different astronomical catalogues by VizieR (http://vizier.u-strasbg.fr) [9] and cross-match service (http://cdsxmatch.u-strasbg.fr/xmatch).